\documentclass[onecolumn,amsmath,amssymb,preprintnumbers,11pt,superscriptaddress,nofootinbib]{revtex4-1}
\usepackage{subfig}
\usepackage{color}
\usepackage{amsmath}
\usepackage{amsfonts}
\usepackage{verbatim}
\usepackage{amssymb}
\usepackage{graphicx}
\usepackage{soul}
\usepackage{mathrsfs}
\providecommand{\U}[1]{\protect\rule{.1in}{.1in}}
\providecommand{\U}[1]{\protect\rule{.1in}{.1in}}

\begin{document}

\allowdisplaybreaks
\begin{titlepage}

\title{$^{}$ \\ Kerr-NUT-de Sitter as an Inhomogeneous Non-Singular Bouncing Cosmology}
\author{Andr\'{e}s Anabal\'{o}n}
\email{andres.anabalon@uai.cl}
\affiliation{Departamento de Ciencias, Facultad de Artes Liberales,
Universidad Adolfo Ib\'{a}$\tilde{n}$ez, Avenida Padre Hurtado 750, Vi$\tilde{n}$a del Mar, Chile}
\author{Sebastian F. Bramberger}
\email{sebastian.bramberger@aei.mpg.de}
\affiliation{Max--Planck--Institute for Gravitational Physics (Albert--Einstein--Institute), Am M\"{u}hlenberg 1, 14476 Potsdam, Germany}
\author{Jean-Luc Lehners}
\email{jlehners@aei.mpg.de}
\affiliation{Max--Planck--Institute for Gravitational Physics (Albert--Einstein--Institute),  Am M\"{u}hlenberg 1, 14476 Potsdam, Germany}

\begin{abstract}
\vspace{.5cm}
\noindent 
We present exact non-singular bounce solutions of general relativity in the presence of a positive cosmological constant and an electromagnetic field, without any exotic matter. The solutions are distinguished by being spatially inhomogeneous in one direction, while they can also contain non-trivial electromagnetic field lines. The inhomogeneity may be substantial, for instance there can be one bounce in one region of the universe, and two bounces elsewhere. Since the bounces are followed by a phase of accelerated expansion, the metrics described here also permit the study of (geodesically complete) models of inflation with inhomogeneous ``initial'' conditions. Our solutions admit two Killing vectors, and may be re-interpreted as the pathology-free interior regions of Kerr-de Sitter black holes with non-trivial NUT charge. Remarkably enough, within this cosmological context the NUT parameter does not introduce any string singularity nor closed timelike curves but renders the geometry everywhere regular, eliminating the big bang singularity by means of a bounce.
\end{abstract}

\maketitle
\end{titlepage}
\tableofcontents

\section{Introduction}

A hundred years ago, the discovery of the expansion of our universe brought with it a complete paradigm shift, in that it implied that our universe is not unchanging, but evolving. The most puzzling consequence of the expansion has been the realisation that, in our past, there must have been a phase of enormous density, called the big bang, which currently represents a true frontier of our knowledge. In the context of general relativity, which provides the framework for interpreting the expansion in the first place, the singularity theorems of Penrose and Hawking \cite{Hawking:1969sw} imply that under rather general conditions the big bang must have been a curvature singularity, at which point general relativity itself breaks down. One may then expect quantum gravitational effects to become of preeminent importance, offering a way to describe the emergence of spacetime out of a quantum state (possibilities that have been put forward include for instance string gas cosmology \cite{Brandenberger:1988aj} or the no-boundary state \cite{Hartle:1983ai}). But there do exist exceptions to the singularity theorems, which may allow for a classical resolution of the big bang in the form of a non-singular bounce. Much studied in recent times have been exotic matter models that allow for violations of the null energy condition while being carefully constructed to avoid a myriad of potential pathologies, such as ghosts, gradient instabilities and causality violations, see e.g. \cite{Qiu:2011cy,Easson:2011zy,Koehn:2013upa,Ijjas:2016tpn,Magueijo:2012ug,Farnsworth:2017wzr}. However, there exists a simple manner in which the singularity theorems may also be avoided, namely by having a spatially closed universe and matter violating only the strong energy condition \cite{Hawking:1967ju}. This is in no way exotic, as dark energy is known to have a pressure that is equal to minus its energy density to better than $10$ percent accuracy \cite{Scolnic:2017caz}, in clear violation of the strong energy condition.  Moreover, although the spatial sections of our current universe are measured to be nearly flat, this is not inconsistent with the early universe having had a significant positive spatial curvature, as long as there is a mechanism that can dissipate the curvature at later times. Inflation is one possibility for such a mechanism, and we will further comment on this below.

Examples of non-singular yet anisotropic bounces in the presence of a cosmological constant were recently studied analytically in \cite{Anabalon:2018rzq}, and numerically in \cite{Bramberger:2019zez} (using the full Bianchi IX metric, and including a scalar field), where the link of such solutions with the singular and chaotic mixmaster/BKL approach to a big bang were pointed out. Here will extend these studies to inhomogeneous solutions, which moreover can contain electromagnetic fields. Not only do these solutions arise in the presence of known matter types, but they also have the great advantage of being exact solutions of general relativity. The solutions possess two Killing vectors, and we will describe their link with the Taub-NUT spacetime \cite{Taub:1950ez,Newman:1963yy}, with black holes with NUT charge \cite{Carter:1968ks}, and more generally how they fit into the Pleba\'{n}ski-Demia\'{n}ksi (PD) class of solutions of the Einstein-Maxwell system \cite{Plebanski:1976gy}. The trouble with such large classes of solutions as the PD class is that the physics they describe remains obscure until one focusses on specific examples and specifies the topology of the spacetime. Thus, even though the solutions that we explore have been known locally for a while, the realisation that the PD class (which is usually regarded as a pathological generalisation of black hole metrics) contains physically sensible inhomogeneous and anisotropic non-singular bounce solutions is to our knowledge novel. Indeed, although less common these days, new interpretations to well-known solutions of the Einstein equations have already occurred several times, see for instance \cite{Finkelstein:1958zz,Kruskal:1959vx,Kinnersley:1970zw}.

We could start by giving the final form of the solutions, but because of the reasons just stated we find it more illuminating to build up the full solution by starting from simple examples, and increasing the complexity step by step. This is useful in appreciating the physical significance of the solutions, and points the way towards a number of applications that we will briefly mention. Clearly, one application of the bounce solutions is as a non-singular replacement of the big bang. But the fact that they may contain electromagnetic field lines also means that they could be useful in addressing the observational problem of early magnetic fields \cite{Subramanian:2015lua}. Finally, because the bounces automatically lead into a phase of accelerated expansion, our solutions may be useful in characterising inflationary models in the presence of anisotropies and inhomogeneities.


\section{Anisotropic Bounces}

The simplest example of a metric that admits a non-singular bounce is pure de Sitter space written in closed coordinates. The metric is then given by \footnote{Indeed, in the case of pure de Sitter spacetime the bounce is an artifact of the coordinates. The de Sitter spacetime is completely homogenous and isotropic. Namely, all the points on the manifold can be reached by means of a isometry. Hence, the location of the bounce is coordinate dependent.} 
\begin{align}
ds^2 = -dt^2 + \ell \cosh\left(\frac{t}{\ell}\right)^2 d\Omega_3^2\,,
\end{align} 
and it solves Einstein's equations in the presence of a cosmological constant $\Lambda=3\ell^{-2}$,%
\begin{align}
R_{\mu\nu}=\frac{3}{\ell^{2}}g_{\mu\nu}\text{ .}%
\end{align}
The spatial sections of the metric above consist of 3-spheres with line element $d\Omega_3^2$, and a first generalisation is to deform these spheres. Let us therefore consider the metric
\begin{equation}
ds^{2}=-\frac{dt^{2}}{N(t)}+g(t)\sigma_{1}^{2}+h(t)\sigma_{2}^{2}%
+f(t)\sigma_{3}^{2}\text{ ,}\label{m1}%
\end{equation}
where the differential forms on a three-sphere are given by
\begin{align}
\sigma_{1} &  =\cos(\psi)d\theta + \sin(\psi)\sin(\theta)d\phi \text{ ,}\\
\sigma_{2} &  =\sin(\psi)d\theta -\cos(\psi)\sin(\theta)d\phi\text{ ,}\\
\sigma_{3} &  =d\psi+\cos(\theta)d\phi\text{ ,}%
\end{align}
with the coordinate ranges $0 \leq \theta \leq \pi, \, 0 \leq \phi \leq 2\pi, \, 0 \leq \psi \leq 4\pi.$ This metric is known as the Bianchi IX spacetime, and it reduces to the closed Friedmann-Lema\^{\i}tre-Robertson-Walker (FLRW) metric when all scale factors are equal. There is an exact analytic cosmology satisfying the Einstein equations with $h(t)=g(t)$, $N(t)=\frac{\sigma^{2}}{4\ell^{4}}f(t)$ and
\begin{align}
f(t) &  =\frac{4\ell^{2}}{\sigma^{2}}\frac{t^{4}+(6-\sigma)t^{2}+\mu
t+\sigma-3}{t^{2}+1}\text{ ,}\\
g(t) &  =\frac{\ell^{2}}{\sigma}(t^{2}+1)\text{ .}%
\end{align}
The function $f(t)$ never vanishes provided%
\begin{equation}
12>\sigma>3\text{ ,}\qquad\left\vert \mu\right\vert <\frac{2}{3\sqrt{3}}%
(12-\sigma)\sqrt{\sigma-3}\text{ ,}%
\end{equation}
and with these inequalities satisfied the solution describes a non-singular bounce. The parameter $\mu$ may be interpreted as the time asymmetry of the metric, which asymptotically approaches de Sitter space as $t \to \pm \infty.$ This bouncing spacetime, with two scale factors being equal, belongs to the class of metrics known as biaxial Bianchi IX. It was obtained in \cite{Anabalon:2018rzq} by analytic continuation from a wormhole solution in asymptotic Anti-de Sitter space. Meanwhile, more general anisotropic bounces in the Bianchi IX class were studied numerically in \cite{Bramberger:2019zez}.

\subsection{Adding an Electromagnetic Field} \label{sec:embounce}

In order to include an electromagnetic field we may add a gauge vector of the form
\begin{align}
A = q(t)\sigma_3\,,
\end{align}
bearing in mind the symmetries of the metric. Then we need to solve the familiar Einstein-Maxwell system of equations
\begin{align}
R_{\mu \nu} - \frac{1}{2} g_{\mu \nu} R + \Lambda g_{\mu \nu} &= T_{\mu \nu}\,,  \\
\nabla^\mu F_{\mu \nu} &= 0\,,
\end{align}
where 
\begin{align}
T_{\mu \nu} = F_{\mu \sigma} F_{\nu}^\sigma - \frac{1}{4}g_{\mu \nu} F_{\rho\sigma} F^{\rho\sigma}\,,
\end{align}
and 
\begin{align}
F_{\mu \nu} = \partial_\mu A_{\nu} - \partial_\nu A_{\mu}
\end{align}
is the anti-symmetric field-strength corresponding to $A$. Maxwell's equations for this gauge vector and metric ansatz reduce to a single equation,
\begin{align}
\ddot{q}+\frac{\dot{g}}{g} \dot{q}+4 \frac{\ell ^4}{\sigma ^2} \frac{q}{g^2} &= 0\,,
\end{align}
which admits the solution 
\begin{align}
q(t) &= 2\sqrt{2} \left( \frac{ q_1t}{t^2+1} + \frac{q_2}{t^2 +1} - \frac{q_2}{2} \right)\,,
\end{align}
with two integration constants $q_1, q_2.$ The Einstein equations are solved provided
\begin{align}
g(t) &= \frac{\ell^2}{\sigma}(t^2 + 1) \\
f(t) &= 4 \frac{\ell^2}{\sigma^2} \frac{t^4 + (6-\sigma)t^2 + \mu t + \sigma -3}{t^2 + 1} - 4 \frac{q_1^2+q_2^2}{t^2 + 1}\,.
\end{align}
To eliminate the magnetic monopoles at large $t$ is necessary to set $q_2=0$ -- we will get back to this case later in the paper. Note that there are new constraints on the allowed parameters now, if we want to avoid reaching a singularity near $t=0$. The absence of a curvature singularity now implies the inequalities
\begin{align}
\mid \mu \mid<\frac{\sqrt{2}}{3\sqrt{3}}\frac{ 6\sigma\left( \sqrt{1-X} + 4 \right) - \sigma^2\left( 1 + X + \sqrt{1-X} \right) - 72}{\sqrt{\sigma \left(1 + \sqrt{1-X} \right)-6}}. \label{regularity}
\end{align}
and $\sigma_{+} > \sigma > \sigma_{-}$ with
\begin{align}
\sigma_{\pm} = 3\frac{4(1\pm3)\sqrt{1-X}}{1+X + \sqrt{1-X}}, \qquad X=\frac{12(q_1^2+q_2^2)}{\ell^2} < 1 \,,
\end{align}
where the bound of $\sigma$ can be found by demanding that the numerator of the bound on $\mu$ never vanishes. An interesting feature of this solution is that the gauge field is non-trivial even though there is no singularity in the metric nor in the gauge field, i.e. there is no source. In fact it is the geometry alone that supports the electromagnetic field lines, and we will explore this aspect in more detail below when discussing the inhomogeneous solution. Here we simply note that the gauge potential grows in the approach to the bounce, and decays again as the universe expands, allowing electromagnetic fields to pass through the bounce. 

\section{Inhomogeneous and Anisotropic Bounces}

Extending from the metric considered thus far, the cosmological solution can be generalised as follows,%
\begin{align}
ds^{2}  & =\ell^{2}\left(  t^{2}+(\alpha y+1)^{2}\right)  \left(
-\frac{dt^{2}}{\Delta(t)}+\frac{dy^{2}}{G(y)}\right)  +\frac{4\ell^{2}}{\sigma^{2}%
}\frac{\Delta(t)+\alpha^{2}G(y)}{t^{2}+(\alpha y+1)^{2}}d\psi^{2} \nonumber \\
& +\frac{4\ell^{2}}{\sigma^{2}}\frac{y\left(  \alpha y+2\right)
\Delta(t)-\alpha\left(  t^{2}+1\right)  G(y)}{t^{2}+(\alpha y+1)^{2}}d\psi d\phi \nonumber \\
& +\frac{\ell^{2}}{\sigma^{2}}\frac{\left(  t^{2}+1\right)  ^{2}%
G(y)+y^{2}\left(  \alpha y+2\right)  ^{2}\Delta(t)}{t^{2}+(\alpha y+1)^{2}}d\phi^{2}\,, \label{metricsol}
\end{align}
where
\begin{align}
\Delta(t)  & =t^{4}+(6+\alpha^{2}-\sigma)t^{2}+\mu t+\left(  \sigma-3\right)
\left(  1-\alpha^{2}\right) \,, \\
G(y)  & =\left(  1-y^{2}\right)  \left(  \alpha^{2}y^{2}+4\alpha
y+\sigma\right)\,.
\end{align}
The homogeneous solutions of the previous section are recovered when $\alpha=0$. The range of the new
coordinate is  $\cos\left(  \theta\right)  =y\in\lbrack-1,1]$. The condition $\alpha^{2}<1$
is necessary for regularity of the metric. Indeed, the would be singularity is at $t=0$
and $y=-\frac{1}{\alpha}$, however this region can never be reached as long as
$\alpha^{2}<1$. Once again we have bounds on the anisotropy parameters we are allowed to take. The effect of $\alpha$ on these is essentially a reduction of the parameter space to obtain non-singular solutions. We shall give the bounds on $\mu$ and $\sigma$ below when discussing the charged solution. The uncharged case can be retrieved by setting the charge to zero.

This solution is a new type of everywhere regular bouncing cosmology when the
range of the parameters is such that $\Delta(t)$ never vanishes. When $\Delta(t)$ has zeroes there exist black hole and cosmological horizons and the
solution is Kerr-Taub-NUT-de Sitter with the standard pathological
interpretation of the NUT parameter. We will comment more on this correspondence below. But when the parameters are chosen such that $\Delta(t)$ remains positive throughout, these solutions describe pathology-free non-singular bounce cosmologies. The parameter $\alpha$ determines the amount of inhomogeneity in the $y$ direction.

\subsection{Adding an Electromagnetic Field}

We may once again add an electromagnetic field. The metric retains the same form as in \eqref{metricsol}, though the function $\Delta$ gets augmented by a term,
\begin{align}
\Delta(t)=t^{4}+(6+\alpha^{2}-\sigma)t^{2}+\mu t+\left(  \sigma-3\right)  \left(
1-\alpha^{2}\right)  -\sigma^{2}\ell^{-2}\left(  q_1^{2}+q_2^{2}\right)\,.
\end{align}
Regularity requires again that $\alpha^2<1$. $\Delta(t)$ must remain positive throughout if we want a singularity-free metric. This condition translates into the following requirement for $\mu$
\begin{align}
\mid \mu \mid < -\frac{1}{3\sqrt{6}}\frac{(12(12-4\sigma-\sigma\sqrt{1-\tilde{X}})+2\sigma^2(1+\tilde{X}+\sqrt{1-\tilde{X}})+\alpha^2(\alpha^2-23\sigma+\sigma\sqrt{1-\tilde{X}}+84))}{\sqrt{-6-\alpha^2+\sigma(1+\sqrt{1-\tilde{X}})}}\,, 
\end{align}
while $\sigma$ must reside in the range 
\begin{align}
\sigma_{+} > \sigma > \sigma_{-}\,,
\end{align} 
with
\begin{align}
\sigma_{\pm}=&\frac{3}{4}\frac{16+4(1-\tilde{X})^{1/2}\pm\Xi^{1/2}}{1+(1-\tilde{X})^{1/2}+\tilde{X}}-\frac{1}{4}\frac{(1-\tilde{X})^{1/2}-23}{1+(1-\tilde{X})^{1/2}+\tilde{X}}\alpha^2\,, \\
\Xi=&144(1-\tilde{X})+168\alpha^2-24\alpha^2(1-\tilde{X})^{1/2}-72\alpha^2\tilde{X}+58\alpha^4-6\alpha^4(1-\tilde{X})^{1/2}-\alpha^4\tilde{X} \,, 
\end{align}
and $\tilde{X}$ must satisfy
\begin{align}
 1\geq \tilde{X}=\frac{12(q_1^2+q_2^2)}{\ell^2}-\frac{\alpha^2(48+\alpha^2-14\sigma)}{\sigma^2}\,,
\end{align}
where the last condition yields a bound on the charge. The vector potential is generalised to
\begin{align}
A=\frac{2\sqrt{2}}{\left(  t^{2}+(\alpha y+1)^{2}\right)}
\left(  \left[  \frac{q_2}{2}\left(  1+y^2\alpha^2 - t^2\right)  -q_1t\right]  d\psi - \frac{y}%
{2}\left[  q_2\left(  t^{2}-1-y\alpha\right)  +q_1t\left(  2+\alpha y\right)
\right]  d\phi\right)
\,,
\end{align}
where $q_1$ and $q_2$ are again the integration constants describing the electromagnetic field.

\begin{figure}[h]
\includegraphics[width=0.6\textwidth]{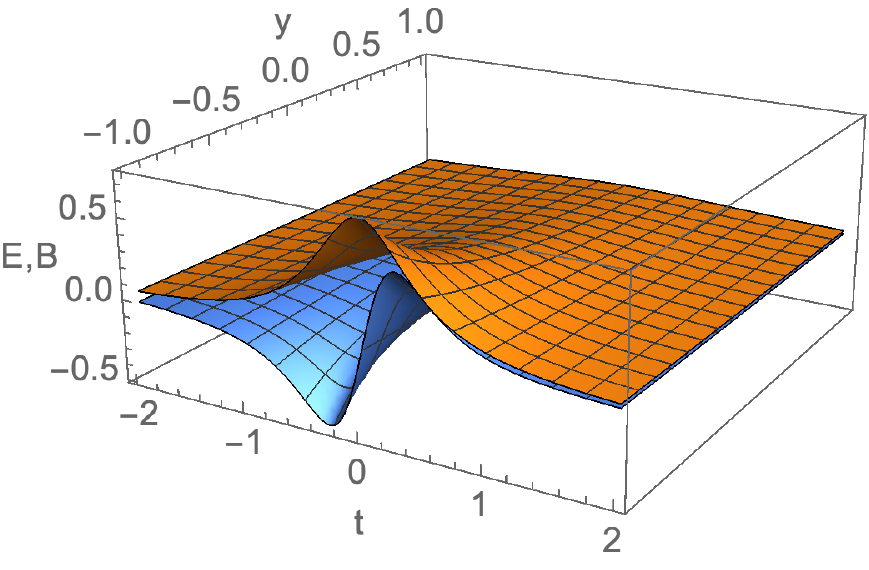}
\caption{The local electric (in orange) and magnetic (in blue) fields for $\sigma=4, \alpha = 1/2$, $\ell^2=3$, $q_1 = -1/10$ and $q_2=0$. For $\alpha > 0,$ the growth of the fields in the approach of the bounce is largest near $y=-1.$} \label{fig:em}
\end{figure}

In order to interpret the gauge potential as giving rise to electric and magnetic fields, we should first shift the description to a local tangent frame. For this we need the vielbeine, which for the metric \eqref{metricsol} are given by
\begin{align}
e^{\bar{0}}_t= \frac{\ell}{\Delta(t)^{1/2}} \left(  t^{2}+(\alpha y+1)^{2}\right)^{1/2} \indent
e^{\bar{1}}_y = \frac{\ell}{G(y)^{1/2}} \left(  t^{2}+(\alpha y+1)^{2}\right)^{1/2} \\
e^{\bar{2}}_\phi = \frac{\ell}{\sigma} y(\alpha y +2) \left( \frac{\Delta(t)}{t^2 + (\alpha y +1)^2} \right)^{1/2}  \indent e^{\bar{2}}_\psi = 2\frac{\ell}{\sigma} \left( \frac{\Delta(t)}{t^2 + (\alpha y +1)^2} \right)^{1/2} \\
e^{\bar{3}}_\phi = \frac{\ell}{\sigma} (t^2 +1) \left( \frac{G(y)}{t^2 + (\alpha y +1)^2} \right)^{1/2} \indent e^{\bar{3}}_\psi = -2 \alpha \frac{\ell}{\sigma} \left( \frac{G(y)}{t^2 + (\alpha y +1)^2} \right)^{1/2}
\end{align}
and all other components are zero. Now we can define the electric and magnetic fields as they would be measured by a local free-falling observer,
\begin{align}
{E}_{\bar{a}} = F^{\bar{0}}{}_{\bar{a}}\,, \qquad
{B}_{\bar{a}} = \frac{1}{2} \epsilon_{\bar{a}bc} {F}^{bc}\,. 
\end{align}
Curiously in local coordinates both the electric and the magnetic fields only point in a single spatial direction,
\begin{align}
E_{\bar{2}} = -\frac{\sqrt{2} \sigma  \left(q_1 \left(t^2-(\alpha  y+1)^2\right)-2 q_2 t (\alpha 
	y+1)\right)}{\ell ^2 \left(t^2+(\alpha  y+1)^2\right)^2}\,, \\
B_{\bar{2}} = \frac{\sqrt{2} \sigma  \left(q_2 \left(t^2-(\alpha  y+1)^2\right)+2 q_1 t (\alpha 
	y+1)\right)}{\ell ^2 \left(t^2+(\alpha  y+1)^2\right)^2}\,,
\end{align}
and all other terms are zero. The Maxwell equations are nevertheless satisfied because the geometry provides the additional terms required. Thus the geometry supports the electric and magnetic fields, which exist without the presence of a source. The general structure of the $E$ and $B$ fields is that they grow in the approach of the bounce, and decay again afterwards. The integration constant $q_1$ corresponds to a time-symmetric electric field and an odd magnetic field (vanishing at $t=0$), while for $q_2$ this correspondence is reversed. An example is shown in Fig. \ref{fig:em}.

\section{A Black Hole - Bounce Correspondence}

In order to appreciate how cosmological and black hole metrics sometimes happen to be related to each other, it is instructive to start with the example of the familiar Schwarzschild black hole metric with mass $M$ \cite{Schwarzschild:1916uq},
\begin{align}
ds^2 = -\left( 1- \frac{2M}{r}\right) dt^2 + \frac{dr^2}{\left( 1- \frac{2M}{r}\right)} + r^2 d\Omega_2^2\,. 
\end{align}
Outside the horizon, $r \geq 2M,$ the spacetime is static with the curvature depending solely on the distance to the horizon. But in the interior of the black hole, $r < 2M,$ the coefficients of $dt^2$ and $dr^2$ switch sign, so that these coordinates exchange their roles -- $t$ becomes a space direction, and $r$ a time direction. Near $r=0,$ the metric can be approximated as
\begin{align}
ds^2 \approx +\frac{2M}{r}dt^2 - \frac{r}{2M}dr^2 + r^2 d\Omega_2^2\,. 
\end{align}
Now we can redefine $r\equiv T^{2/3}$ and call $t\equiv R,$ with the consequence that up to some trivial re-scalings the metric becomes
\begin{align}
ds^2 \approx - dT^2 + T^{-2/3} dR^2 + T^{4/3} d\Omega_2^2\,.
\end{align}
This shows that in the black hole interior the metric is of Kantowski-Sachs type \cite{Kantowski:1966te}, i.e. it has the topology $\mathbb{R}^2 \times S^2.$ Near the centre of the Schwarzschild black hole, at $T=0,$ the metric is of approximate Kasner form with exponents $(-\frac{1}{3},\frac{2}{3},\frac{2}{3}).$ In other words, the interior of the Schwarzschild black hole is a time-dependent contracting universe ending in a big crunch singularity at $T=0.$ From the point of view of classical general relativity, this interior solution is not particularly useful (although one may speculate what the fate of the crunch may end up being in quantum gravity). But for more general black hole metrics, the interior region can be considerably more interesting.

We will be particularly interested in the Kerr-Newman-NUT-deSitter solution in Boyer-Lindquist coordinates \cite{Griffiths:2005qp}, 
\begin{align}
ds^2 = & -\frac{Q}{\rho^2} \left[ dt - \left( a \sin^2\theta + 4 n \sin^2 \frac{1}{2}\theta \right)d\phi \right]^2  + \frac{P}{\rho^2} \sin^2\theta \left[a dt - \left(r^2 + (a + n)^2 \right)d\phi  \right]^2 \nonumber \\ &+ \frac{\rho^2}{Q} dr^2  + \frac{\rho^2}{P} d\theta^2
\end{align} 
with
\begin{align}
\rho^2 &= r^2 + (n + a \cos \theta)^2\,, \label{radial} \\
P &= 1 + \frac{4}{3} \Lambda a n \cos \theta + \frac{1}{3} \Lambda a^2 \cos^2\theta\,, \\
Q &= (a^2 - n^2 + e^2 + g^2) - 2mr + r^2 - \Lambda \left[ (a^2 - n^2) n^2 + (\frac{1}{3}a^2 + 2n^2)r^2 + \frac{1}{3}r^4 \right]\,, \label{Q}
\end{align}
where $m$ is the mass, $e$ and $g$ are the electric and magnetic charges, $n$ is the NUT parameter, $a$ is the spin and $\Lambda$ is the cosmological constant. Horizons are located at zeroes of $Q.$ Meanwhile the corresponding vector potential is given by
\begin{align}
A=\frac{2\sqrt{6}}{\sqrt{\kappa \Lambda}}\frac{1}{\rho^2 n^2}
\left( d\psi \left[  gn (n +a\cos\theta) - enr\right] +\frac{\cos\theta}
{2} d \phi \left[  g\left( n^2r^2 -1 - a\cos\theta \right) +e\left(  2n+a \cos\theta \right)
\right] \right)
\end{align}
From here we make the following coordinate transformations and redefinitions
\begin{align}
r  = t \sqrt{\frac{\ell^2}{\sigma}}\,, \qquad \cos \theta = y\,, \qquad t = \psi\,, \\
a = \alpha \sqrt{\frac{\ell^2}{\sigma}}, \indent n =  \sqrt{\frac{\ell^2}{\sigma}}, \indent m = \frac{1}{2} \mu \ell \sigma^{-3/2}, \indent &\Lambda = \frac{3}{\ell^2}, \indent e = \frac{\ell}{\sigma}q_1, \indent g = \frac{\ell}{\sigma}q_2\,,
\end{align}
which precisely recover the inhomogeneous/anisotropic non-singular bounce solution \eqref{metricsol} we have used above. Horizons would be located at zeroes of $Q,$ but we have chosen parameters and coordinate ranges such that for the bouncing solution $Q<0$ everywhere. This means that one should think of the bouncing cosmology as the smooth joining of the region outside the cosmological horizon with the region inside the event horizon of the Kerr-NUT-de Sitter black hole. The fact that the cosmological constant is positive is in fact crucial for this to be possible, as can be seen from Eq. \eqref{Q}. Note also that a curvature singularity is reached at $\rho=0.$ But from Eq. \eqref{radial} we can see that if the NUT parameter $n$ is larger than the rotation $a,$ then $\rho$ can never be zero. In our notation this translates into the requirement $\alpha^2<1,$ so that we can see that a sufficiently large NUT charge is  required to change the big bang into a non-singular bounce. In the stationary region of the black hole, the NUT charge is considered pathological, as it leads to the appearance of closed timelike curves, but in the interior region it takes on the new role of preventing a singularity. When the cosmological constant is positive this interior region can be extended to a geodesically complete spacetime representing the bouncing cosmologies we discuss here.

Note also that the switch between spacelike and timelike directions means that the mixed time-space component of the metric morphs into a mixed spatial component only, and, together with the definite sign of all metric coefficients, this is the reason why no closed timelike curves can appear in the interior region. Related to this is the fact that $a$ no longer characterises the rotation/angular momentum of the black hole, and in fact comes to parameterise the spatial inhomogeneity $\alpha$ of the bounce. Finally, we note that the mass $m$ of the black hole ends up simply parameterising the time asymmetry $\mu$ of the bouncing solution. Thus there is a complete re-shuffle of the physical significances of the various parameters, the most important one being that the NUT charge $n$ looses its stigma. In the cosmological setting the NUT parameter is controlled by $\sigma$, which measures the amount of anisotropy that the metric has at large times. When $\sigma=4$ the anisotropic cosmology evolves towards the closed FLRW metric with a round sphere.


\section{Examples}

In order to highlight the non-trivial features of the non-singular bounce solutions that we have described so far, it is useful to present a few representative examples. These examples may also point to several directions of research that will be worthwhile exploring in more detail in the future. We will characterise the solutions by looking at the size of the spatial hypersurfaces at fixed times, and at the contributions of the various forms of gravitational and matter energy densities that determine the contraction/expansion history of the solutions. 

\begin{figure}[h]
\includegraphics[width=0.45\textwidth]{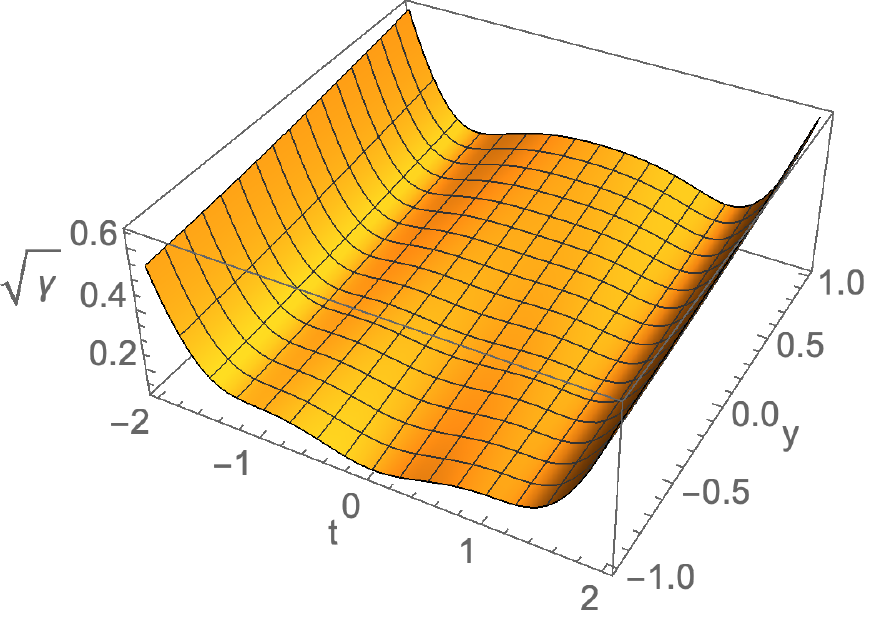}
\includegraphics[width=0.45\textwidth]{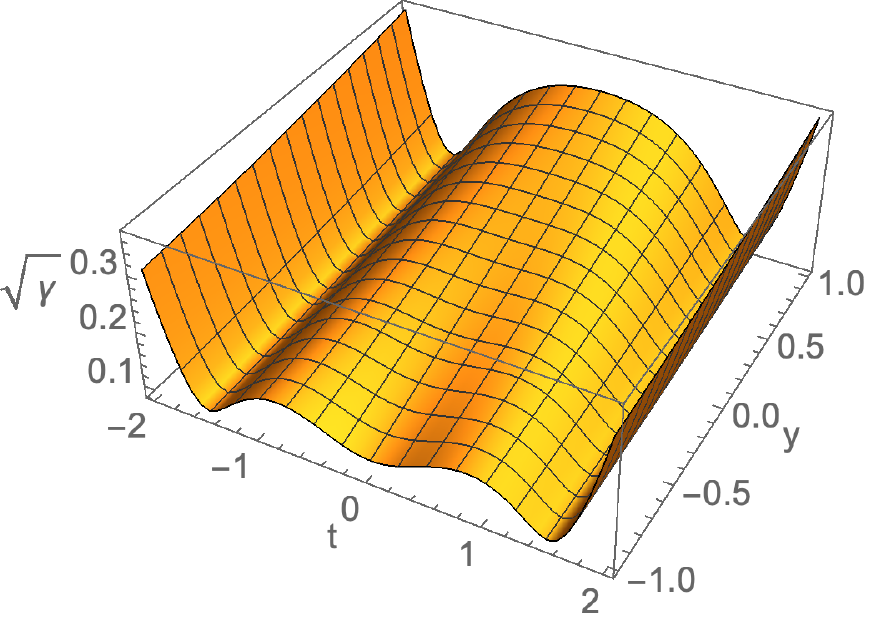}
\caption{Plots of the average local scale factor (cubed) as a function of $y$ and $t.$ The solutions can be significantly inhomogeneous: for instance there can be one bounce near $y=-1$ and two bounces near $y=+1$ (left panel, with $\alpha=1/2, \sigma = 10, q_1=q_2=0$) or two bounces on one side and three on the other (right panel, with $\alpha=1/2, \sigma = 11, q_1=q_2=0$). All our plots have $\ell^2=3, \mu=3.$} \label{fig:bounces}
\end{figure}

The solutions that we are describing are both anisotropic and inhomogeneous. Nevertheless, we can define a local scale factor $A(t,y)$ which averages over the anisotropies, but shows the inhomogeneity and the dependence on time, by making use of the determinant of the metric $\gamma_{ij}$ on constant $t$ slices (the integral of which would yield the volume),
\begin{align}
A(t,y)^3\equiv \sqrt{\gamma} = \frac{2\ell^3}{\sigma^2}\left[ (t^2+(\alpha y + 1)^2)\Delta(t)\right]^{1/2}\,.
\end{align}
This allows us to highlight an interesting feature of the bounces: there are solutions for which the inhomogeneity is so large that the number of bounces a local observer experiences depends on the location in $y,$ see Fig. \ref{fig:bounces}. As the figure shows, there exist solutions where one region of the universe bounces once, while far away regions bounce twice. Likewise, there are solutions containing two or three bounces depending on location. Three bounces is however the maximum possible number, since the equation $\dot{A}=0$ contains five real roots at most, corresponding to three bounces separated by two occurrences of re-collapse. Asymptotically however, as $t \to \pm \infty,$ the metric becomes independent of $\alpha$ and the inhomogeneity is diluted -- hence it is only near the bounce(s) that the inhomogeneity is really pronounced.

Another useful way to characterise the inhomogeneity as well as other features of the solutions is to look at the contributions of the different forms of stress energy: gravitational, vacuum and of electromagnetic type. For this, it is convenient to decompose the metric \eqref{metricsol} into a $3+1$ split \cite{Arnowitt:1962hi},
\begin{align}
g_{00} = - N(t,y)^2\,, \indent g_{0i} = 0\,, \indent g_{ij} = \gamma_{ij}\,, 
\end{align}
where we note that for our metric the shift is equal to zero (here we use Roman letters for spatial indices and Greek ones for spacetime indices). The three dimensional hypersurfaces have extrinsic curvature $K_{ij}$ arising from their embedding into the four-dimensional spacetime,
\begin{align}
K_{ij} = -\frac{1}{2N} \partial_t \gamma_{ij}\,.
\end{align}
Then the time-time component of the Einstein equations, usually referred to as the Friedman equation in a cosmological context, reads
\begin{align}
\frac{1}{2} \left(  K^2 -K_{ij}K^{ij} + {}^{(3)}R \right) = T^0{}_{0} = \frac{3}{\ell^2} + \frac{\sigma ^2}{\ell ^4}\frac{ \left(q_1^2+q_2^2\right)}{ \left[t^2+(\alpha  y+1)^2\right]^2} \,.\label{Friedman}
\end{align}
On the right hand side of Eq. \eqref{Friedman} we have contributions both from the cosmological constant and from the stress-energy of the electromagnetic field. The left hand side, given in terms of the extrinsic curvature and the three-curvature, reads more explicitly
\begin{align}
K^2 - K_{ij}K^{ij} &= \frac{2 t \left( \dot{\Delta} \left(t^2+(\alpha  y+1)^2\right)-t \Delta - \alpha ^2 t G\right)}{\ell ^2 \left(t^2+(\alpha  y+1)^2\right)^3}\,, \label{extrinsic} \\
{}^{(3)}R &= -\frac{2 (\alpha  y+1)^2 (\Delta + \alpha^2 G)+\left(t^2+(\alpha  y+1)^2\right) \left(G'' \left(t^2+(\alpha  y+1)^2\right)-2 \alpha  (\alpha  y+1) G'\right)}{\ell ^2 \left(t^2+(\alpha  y+1)^2\right)^3}\,.
\end{align}
In a FLRW context the extrinsic curvature term \eqref{extrinsic} would simply have been $3H^2$ (where $H$ denotes the Hubble rate), while the spatial curvature term would have been $\frac{3k}{a^2}$ for spatial slices that are closed ($k=1$), flat ($k=0$) or open ($k=-1$). In such a FLRW context a positive curvature term is needed in order to obtain a non-singular bounce. Meanwhile, in the present inhomogeneous context, all these terms, apart from the cosmological constant term, can have a strong spatial and temporal dependence.

\begin{figure}[h]
\includegraphics[width=0.54\textwidth]{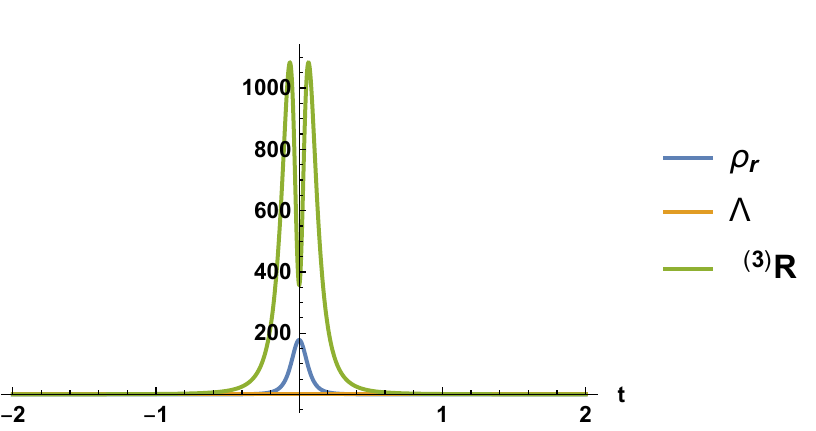}
\includegraphics[width=0.44\textwidth]{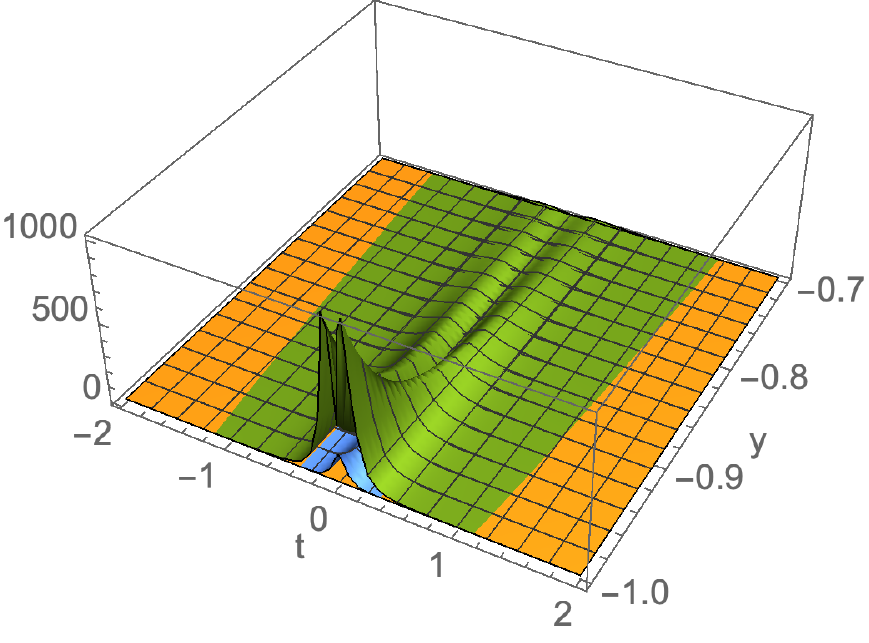}
\caption{An example with large spatial curvature. In blue is shown the energy density of the electromagnetic field, in orange that of the cosmological constant (equal to $\Lambda=1$), while the green curve/surface shows the $3$-curvature. The parameter values are $\sigma=8, \alpha = 9/10, q_1=1/20, q_2=0.$ Left panel: $y=-1.$ Right panel: $-1 \leq y \leq -0.7.$} \label{fig:largecurv}
\end{figure}

As a first example, consider Fig. \ref{fig:largecurv}. This provides an example of a highly inhomogeneous solution, with $\alpha = 9/10$. We are plotting various contributions to the Friedman equation: in blue, the stress-energy from the electromagnetic field, in orange that of the cosmological constant (set to $\Lambda=1$ here) and in green we are showing ${}^{(3)}R.$ Both the electromagnetic energy density and the $3$-curvature are growing towards the bounce, and then decaying again. Near $y=-1$ the growth is by far the strongest, and in this region the $3$-curvature can become a full three orders of magnitude larger in magnitude than the cosmological constant. The right panel shows that this growth is far less pronounced at larger $y.$ 

This solution is also interesting in the context of inflation. An unresolved open problem of all inflationary models is how to explain the initial conditions that are required for inflation to begin \footnote{Quantum cosmology may offer a setting where this question can be addressed. In particular, this is the aim of the no-boundary \cite{Hartle:1983ai} and tunnelling proposals \cite{Vilenkin:1982de}. For recent progress in this direction, see \cite{DiTucci:2019dji}, and for implications in the context of bouncing cosmologies see \cite{Battarra:2014xoa,Battarra:2014kga,Bramberger:2017cgf}.}. But it remains an open issue in and of itself to understand in general what the range of allowable initial conditions is (for recent work see e.g. \cite{Creminelli:2019pdh}). An intuitive expectation would be to require a Hubble sized region to be roughly homogeneous and isotropic, with inflationary potential energy dominating over the kinetic energy. Recently, numerical studies have largely confirmed these expectations, but have also indicated that a larger inhomogeneity may in fact be tolerable (while still assuming the inflaton kinetic energy to be very small) \cite{East:2015ggf,Clough:2016ymm}. Our explicit analytic bounce solutions are interesting in this regard, as they all link to a phase of accelerated/inflationary expansion, albeit one induced by a cosmological constant, where the issues with kinetic energy do not arise. Our solutions demonstrate that the inhomogeneity can indeed be surprisingly large, while still allowing accelerated expansion to take place afterwards. Nevertheless, one should note that in the present case the regions of large curvature are surrounded by regions of small curvature at larger $y,$ so that it may also be the case that these low curvature regions are pulling the large curvature regions along into the ensuing phase of accelerated expansion. It would certainly be interesting to study these questions numerically for initial conditions that are obtained as deformations of the exact solutions presented here, to verify the robustness of the comments above.  

\begin{figure}[h]
\includegraphics[width=0.54\textwidth]{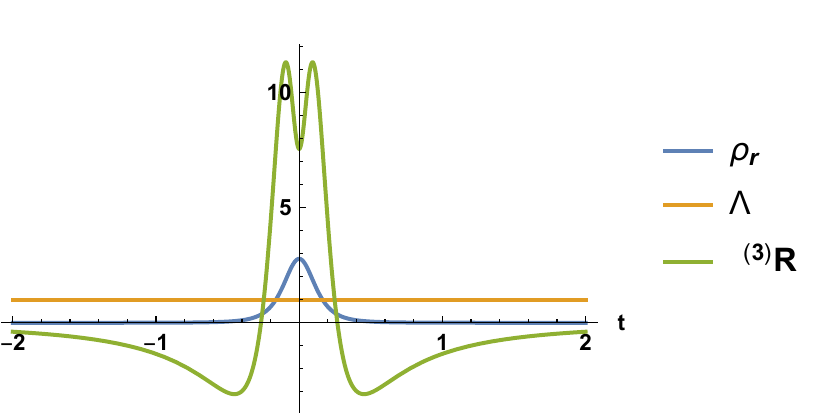}
\includegraphics[width=0.44\textwidth]{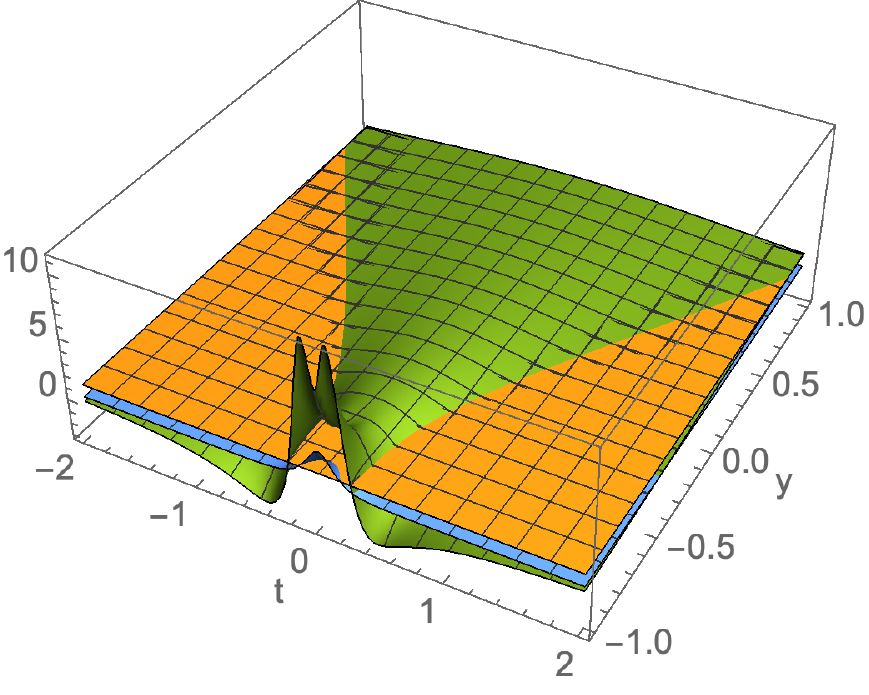}
\caption{An example where the spatial curvature changes sign in some regions, just before and after the bounce, which occurs at $t=0$. In blue is shown the energy density of the electromagnetic field, in orange that of the cosmological constant (equal to $\Lambda=1$), while the green curve/surface shows the $3$-curvature. The parameter values are $\sigma=4, \alpha = 4/5, q_1=1/20, q_2=0.$ Left panel: $y=-1.$ Right panel: $-1 \leq y \leq +1.$} \label{fig:negcurv}
\end{figure}

Another example of interest is presented in Fig. \ref{fig:negcurv}. Here a different, though equally surprising effect takes place. As discussed in the introduction, it is the combination of vacuum energy and positive $3$-curvature that allows the singularity theorems to be evaded. Thus we know that at the bounce the $3$-curvature is necessarily positive. However, for a significant range of parameters, the $3$-curvature switches sign and becomes negative right before/after the bounce, again in the region of the largest inhomogeneity, near $y=-1$. This is interesting again in the context of ``initial'' conditions, in particular regarding the flatness problem \cite{Dicke:1900mn}. From the fact that current observations provide a stringent upper bound on the homogeneous spatial curvature today, we can infer that at the onset of the hot big bang phase the relative importance of the $3$-curvature must have been extremely tiny. Considering that non-singular bounces (without exotic matter that can violate the null energy condition) require a significant positive spatial curvature then seems to be in direct conflict with observations, unless there exists a mechanism that dilutes this curvature after the bounce. Of course, inflation could potentially provide such a mechanism \cite{Guth:1980zm}. But here we see that the case against pure curvature-induced bounces is perhaps less watertight than assumed so far: the fact that the $3$-curvature can change sign right after the bounce also implies that it will vanish or nearly vanish in some regions. It would be a strong use of the anthropic principle to simply postulate that we might live in such a region, and we do not want to pursue this line of reasoning here. However, we simply wish to point out that it might be interesting to investigate this question further, and to see under what conditions the dynamics might cause large regions of the universe to become flat or nearly flat in the aftermath of a non-singular bounce.


\section{Discussion}

Exact solutions of general relativity, in the presence of well-understood matter sources, have played a leading role not only in understanding the structure of relativity itself, but also in understanding its physical consequences for the universe. The most obvious examples that come to mind are the Schwarzschild solution describing the simplest black holes, and the Friedman-Lema\^{i}tre-Robertson-Walker solutions describing the evolution of the universe on the largest scales. In the present paper we have morphed a generalised exact black hole solution, namely the Kerr-NUT-de Sitter solution, into a cosmological solution, by focussing on the matching of the interior region of the black hole to the asymptotic region by eliminating the event horizon. This solution, which exists in the presence of a cosmological constant and (optionally) an electromagnetic field, is distinguished by being both anisotropic and inhomogeneous while describing a non-singular bouncing universe.

Could this solution describe the interior of actual black holes? And could such a non-singular bounce lead into a new expanding universe on the ``other'' side of the black hole, as has been suggested in some scenarios of cosmic evolution \cite{Smolin:1997rs}? Unfortunately this seems unlikely, as the black holes in question are known to lead to closed timelike curves outside of their horizon due to the presence of the NUT charge, implying that this class of black holes is unlikely to be physically realistic. However, on the inside of these black holes, the various parameters describing the solution take on entirely different meanings, and it is precisely the NUT parameter that pushes the would-be big bang singularity out into an unphysical coordinate range, thus rendering the solution everywhere regular. Meanwhile, the rotation parameter of the black hole ends up describing the inhomogeneity of the bouncing universe solution. The end result is that the bouncing solution is entirely non-pathological.

Could the bouncing solution describe our universe? This remains too early to tell. We do however foresee a number of applications of this solution: for instance, as an exact inhomogeneous solution, it may well have applications in terms of understanding the averaging problem in cosmology better \cite{Buchert:1995fz}. And since the bounce is followed by a period of accelerated expansion, these solutions may be useful in understanding the initial conditions required for phases of accelerated expansion, i.e. for inflation. Indeed, the issue of how much inhomogeneity is tolerable while still allowing for inflation to get started remains incompletely understood. Most of the recent work in this direction has focussed on numerical techniques, but exact solutions certainly have a role to play in this context, not only to check the accuracy of numerical codes, but also to understand and perhaps uncover qualitatively new effects. 

From the purely classical point of view the existence of these solutions is quite satisfactory. Indeed, given a classical field theory it is necessary to find solutions that represent the physics one is trying to describe. Therefore, any singularity of the field should be ruled out in the description of the origin of the universe. The existence of a singularity implies that the solution is not a good model of the region of interest. Indeed, it is likely that the final understanding of the origin of the universe shall require a quantum theory of gravity. However, it is natural to expect perturbation theory around a state where observables are infinite to be ill-defined. Hence, if one expects the existence of a regime where the putative theory yields quantum ``corrections'' it is necessary to have at hand configurations where the gravitational field is bounded. 

It remains an open question then how such non-singular bounces might fit into a more complete cosmological model \cite{Lehners:2008vx}. However, the mere fact that they can occur in the presence of known matter sources already motivates their study. Somewhere, sometime, the conditions may have been (or will be) just right for them to actually take place. But, perhaps most intriguingly, they display some features that seem worth further exploration: they support electromagnetic fields simply due to their intricate geometry, and these fields grow in the approach to the bounce. It would be interesting to see if there can be any connection with the magnetic fields that are speculated to have been present already in the early universe. Also, the fact that the $3$-curvature can vary widely from place to place, and even switch sign in some regions, offers new avenues of inquiry. The usual objections to non-singular curvature-induced bounces, namely that they require highly homogeneous initial conditions, and that the required spatial curvature is eventually at odds with current bounds on the curvature, though not evaded are at the very least relativised by the existence of these solutions. After all, in an inhomogeneous universe not all regions are the same, and some neighbourhoods may be much more interesting than others.

\vspace{1cm}

\begin{center}
{\it Acknowledgments}
\end{center}

\noindent The research of AA is supported in part by Fondecyt Grant 1181047, 1170279 and 1161418. 
SFB and JLL gratefully acknowledge the support of the European Research Council in the form of the ERC Consolidator Grant CoG 772295 ``Qosmology''. SFB would like to thank the Studienstiftung des Deutschen Volkes for their support.

\bibliography{InhomogeneousBounce}


\end{document}